\documentclass[aps,prd,twocolumn,groupedaddress, showpacs]{revtex4} 

\usepackage{graphicx}
\usepackage{amsmath,amssymb} 

\begin{document}

\title{Density perturbations in braneworld cosmology and primordial black holes}

\author{Edgar Bugaev}
\email[e-mail: ]{bugaev@pcbai10.inr.ruhep.ru}

\author{Peter Klimai}
\email[e-mail: ]{pklimai@gmail.com}
\affiliation{Institute for Nuclear Research, Russian Academy of
Sciences, 60th October Anniversary Prospect 7a, 117312 Moscow, Russia}




\begin{abstract}

We study, by numerical methods, the time evolution of scalar perturbations
in radiation era of Randall-Sundrum braneworld cosmology. Our results confirm an existence
of the enhancement of perturbation amplitudes (near horizon crossing), discovered recently.
We suggest the approximate solution of equations of the perturbation theory in the high
energy regime, which predicts that the enhancement factor is asymptotically constant, as 
a function of scale. We discuss the application of this result for the problem of primordial
black hole production in braneworld cosmology.

\end{abstract}


\maketitle

\section{Introduction}

During last decade, braneworld cosmological scenarios, in which our 4D Universe
is realized as a hypersurface embedded in a higher-dimensional spacetime
have attracted much attention.
In first scenarios of this kind, suggested as early as in 1980's \cite{Akama:1982jy, Rubakov:1983bb},
it had been shown that matter fields can be confined to a
field-theoretical domain wall (topological defect) in a world with
non-compact extra dimensions.
The progress in string theory in subsequent years, especially the discovery of D-branes, has revived
interest to the idea of braneworlds.
In general, the string theory is quite promising, it
may provide an unified description of gauge interactions and gravity.
In the present context, it is most important that
it predicts the existence of $p$-branes, ($p+1$)-dimensional sub-manifolds of the
10 (or 11) - dimensional spacetime on which open strings end.
Gauge particles and fermions which correspond to string end points can only move
along these $p$-branes, while gravitons can propagate in the full spacetime (``bulk'').
It is tempting to assume that our ($3+1$)-dimensional spacetime is such a 3-brane.
If only gravity can probe the bulk, the extra dimensions can be very large (in comparison
with the smallest length scale tested, so far, in particle physics, $\sim 10^{-16}\;$cm).
It had been assumed in \cite{ArkaniHamed:1998nn}, that the extra dimensions are compact,
in analogy with the old Kaluza-Klein (KK) picture \cite{KK}. Slightly later, in works by
Randall and Sundrum \cite{Randall:1999ee, Randall:1999vf}, it was pointed out that this
condition is not necessary and the extra dimension may be even non-compact.

The Randall-Sundrum (RS) model is of particular interest due to its
relative simplicity, in spite of the fact that it includes nontrivial gravitational
dynamics. In the RS2 model \cite{Randall:1999vf} a single brane is embedded in a
anti - de Sitter (AdS) bulk and, although the 5th dimension extends infinitely,
the warped structure of the bulk geometry (i.e., the curvature of the bulk spacetime)
leads to a recovery of the standard General Relativity (GR) on the brane at scales larger
than the bulk curvature scale $\ell$.  In particular, Newton's law is recovered at
large distances and the Friedmann's equation for the evolution of the Universe is obtained
at low energy.

At high energies, i.e., in the very early Universe, the Friedmann equation differs
substantially from GR by a correction term which is proportional to
$\rho/\sigma$, where $\rho$ is the density of brane matter, and $\sigma$ is the brane tension.
This term leads to a faster Hubble expansion at high energies. Inflationary
expansion of the Universe is also modified in brane cosmology: the evolution
of the inflaton field is more strongly damped, and the brane Universe inflates at
much faster rate than what is expected from standard cosmology.
Another important effect at high energies is the excitation of KK-modes which escape
from our brane into the 5D bulk, leading, in particular, to the suppression of the
power spectrum of inflationary gravitational wave background.

Cosmological perturbation theory in braneworld cosmology also has some distinct features
\cite{Mukohyama:2000ui, Kodama:2000fa, vandeBruck:2000ju, Langlois:2000ph, Koyama:2000cc, Deffayet:2002fn}.
The equations of the perturbation theory contain high-energy corrections ($\sim\rho/\sigma$)
similar to those in the Friedmann equation and, in addition, the correction terms
arising from the fluctuations of the bulk geometry. Perturbations on brane, e.g.,
the scalar perturbations (which we are interested in) are coupled with the bulk
perturbations. Technically, in a case of the scalar perturbations and AdS bulk, the
problem is reduced to the solution of a system of equations for the density contrast variable
and the so-called master variable (it appears that all quantities describing the
bulk perturbations are written in terms of this
variable \cite{Mukohyama:2000ui, Mukohyama:2001yp, Kodama:2000fa}).

In the context of braneworld models, a question about existence and evolution laws
of the higher-dimensional black holes is very interesting and important.
In a model with the 5th large extra dimension, a physically meaningful
black hole solution is the 5D-Schwarzschild \cite{Tangherlini:1963bw, Myers:1986un},
if the horizon size is sufficiently small compared with an effective size of the extra
dimension. Really, it is natural to assume that primordial braneworld black holes formed in the early Universe
with a horizon size $r_s \ll \ell$ would be described by a 5D Schwarzschild metric because
in this case the AdS curvature has very little effect on the geometry. Numerical calculations
support the existence of static solutions for such small $r_s$ \cite{Kudoh:2003xz}.
However, the results of these calculations
cannot be extrapolated to the case $r_s\sim \ell$.

Unfortunately, an exact solution
representing a localized and stable black hole is known only in 4D braneworld
model \cite{Emparan:1999wa}, whereas the corresponding solution in the 5D braneworld model
has not been found. The process of the gravitational collapse on the brane is very complicate,
due to, in particular, gravitational interaction between the brane and the bulk
(see, e.g., \cite{Maartens:2010ar}).
Even in the simplest case of RS-type brane, and Oppenheimer-Snyder (OS) - like collapse,
braneworld gravity introduces important new features in the black hole formation process
(the high energy- and KK-corrections to the field equations of GR,
i.e., the same corrections which affect the expansion of the early Universe, are
also efficient here).
These features lead to a non-static exterior of the black hole \cite{Bruni:2001fd}
in the case of the OS-collapse. Moreover, there are arguments
\cite{Tanaka:2002rb, Emparan:2002px} based on AdS/CFT-correspondence, that the non-static
behavior exists also in a general collapse.
If, really, the black hole solutions in braneworld scenarios, for a black hole larger than AdS radius,
are quite different from those in 4D GR (i.e., if, as authors of
\cite{Tanaka:2002rb, Emparan:2002px} argue, these solutions are necessarily non-static and predict
short lifetime of large black holes due to the strongly enhanced evaporation),
there is an unique chance to probe the extra
dimension by astronomical observations of massive black holes.

Predictions for an evolution of the small ($r_s\ll \ell)$ black holes
are less dramatic (and less speculative). The differences from the 4D case
are reduced to a larger probability of accretion, in the high energy regime (due to
the fact that in this regime the radiation density is proportional to $t^{-1}$ rather
than $t^{-2}$) and to a relative increase of the
primordial black hole (PBH) lifetime, for a given initial mass.
In particular, initial mass of PBHs evaporating today can be $10^9 - 10^{10}\;$g rather
than $\approx 10^{15}\;$g as predicted by GR.

For the PBHs having small masses, there are astrophysical constraints on their
abundance, based, e.g., on studies of extragalactic photon and neutrino backgrounds.
These constraints give, as usual, the information about primordial density
perturbations (we assume that PBHs form from these perturbations).
For an extraction of this information one must know the evolution
of these perturbations in radiation era. In the recent work by Cardoso {\it et al}
\cite{Cardoso:2007zh} it had been shown that the density perturbations with short
wavelengths are amplified during horizon re-entry. The magnitude of this enhancement
depends, clearly, on a scale of the density perturbations.
The smaller is the scale, the earlier the perturbation crosses horizon, and, if comoving
wave number $k$ is larger than some critical value $k_c$, this crossing happens at high energy
regime.
The straightforward calculation of the enhancement factor, in the region of scales which are relevant for
PBHs with small masses, evaporating near today, is quite difficult, even numerically,
due to a very complicate machinery of cosmological perturbation theory in
braneworld cosmology.

In the present paper we study the dependence of the enhancement factor on
the comoving size of the density perturbations. We carried out detailed numerical calculations
of gauge invariant amplitudes of curvature perturbations as functions of the scale factor
and the corresponding enhancement factors.
We found the approximate solution (of the equations for perturbation amplitudes), describing the
time evolution of the amplitudes near horizon crossing. According to this solution, the magnitude
of the enhancement factor doesn't depend, in the high energy region, on the comoving scale.
Using this conclusion it is possible to calculate the enhanced perturbation amplitudes
for arbitrarily small scale.

The plan of the paper is as follows. In the second section the equations of
perturbation theory in RS2 braneworld cosmology
which are necessary for curvature perturbation calculations are given.
In the third section, the approximate solution of these equations in the
high energy limit is suggested. In the fourth section the main relations
characterizing the PBH evolution in the RS2 braneworld are briefly
reviewed. The scheme used in the numerical calculations is presented in Sec. \ref{sec-scheme}.
The results of the paper and conclusions are summarized in the last
section.

\section{Scalar perturbations in RS2 model}
\label{SecSP}

\subsection{Braneworld cosmology in RS2 model}

More than ten years ago, in works \cite{Kraus:1999it, Ida:1999ui, Binetruy:1999hy, Mukohyama:1999qx},
exact cosmological solutions in the braneworld had been obtained. It was shown also
\cite{Mukohyama:1999wi}, for the case when the bulk is empty, that five-dimensional
geometry of all these cosmological solutions is the well-known \cite{Birmingham:1998nr}
Schwarzschild-AdS (Sch-AdS) spacetime (i.e., the spacetime with 5D black hole geometry), having
the metric
\begin{equation}
\label{ds5-h}
  {}^{(5)\!}ds^2= - h(r) d\tau^2 + \frac{dr^2}{h(r)} + r^2 d\Sigma_K^2.
\end{equation}
Here, $d\Sigma_K^2$ is a metric of a unit 3D sphere, plane or hyperboloid (for $K=+1, 0, -1$,
respectively),
\begin{equation} \label{hr}
  h(r) = K + \frac{r^2}{\ell^2} - \frac{M}{r^2},
\end{equation}
$K$ is the curvature of the horizon, $M$ is the mass parameter of the black hole
at $r=0$, $\ell$ is the AdS
curvature radius. The most natural physical interpretation is that a cosmologically evolving
brane is moving in this spacetime, while for an observer on the brane this motion will
be seen as an expansion of the universe. If the brane trajectory is given by equations
$\tau_b=T(t)$, $r_b=a(t)$, where $t$ is the proper time of the brane, the induced metric on
the brane becomes
\begin{equation}
  {}^{(4)\!}ds^2= -  d t^2 + a^2(t) d\Sigma_K^2,
\end{equation}
which is the metric of the Friedmann-Lemaitre-Robertson-Walker (FLRW) spacetime.

The parameter $M$ is unknown, but the value of it can not be too large,
for the braneworld scenario to be consistent, e.g., with nucleosynthesis
data \cite{Binetruy:1999hy}. We suppose that $M=0$ and shall consider below
only this particular case. Further, we shall consider the spatially flat brane only,
i.e., $K=0$. Introducing a new spatial coordinate $z$ by relation $z=\ell/r$, the
metric (\ref{ds5-h}) with $M=0, K=0$ becomes conformally flat,
\begin{equation}
\label{eq4}
  {}^{(5)\!}ds^2= 
   {\ell^2\over z^2} \left(-d\tau^2 + dz^2 + \delta_{ij} dx^i dx^j  \right).
\end{equation}
On the brane, the connection of $\tau$ and $t$ is given by \cite{Mukohyama:2001yp}:
\begin{equation}
\tau_b=T(t), \;\; \dot T = \frac{1}{a} \sqrt{1+\ell^2 \left( \frac{\dot a}{a}\right)^2 } \; ,
\end{equation}
and a $z$-coordinate of the brane is $z_b=\ell/r_b=\ell/a$.

The main hypothesis of any braneworld model is that the string theory predicts Einstein
gravity in the bulk, i.e., the equation
\begin{equation}
  G_{AB} = \kappa_5^2 {\cal T}_{AB}
\end{equation}
takes place. In our case, the bulk energy-momentum tensor has the form
\begin{equation}
  {\cal T}_{AB} = -\frac{\Lambda_5}{\kappa_5^2} G_{AB} + \delta_A^\mu \delta_B^\nu S_{\mu\nu} \delta(y-y_b),
\end{equation}
where $\Lambda_5$ is the bulk cosmological constant, $\kappa_5$ is a 5D gravitational
coupling constant, $S_{\mu\nu}$ is an effective energy-momentum tensor for the brane,
$y_b$ is the brane position (transverse coordinate of the brane). In
the Gaussian normal (GN) system, $y_b=0$.
The tensor $S_{\mu\nu}$ consists of a brane tension term and the matter energy-momentum
tensor $T_{\mu\nu}$,
\begin{equation}
 S_{\mu\nu} = \sigma g_{\mu\nu} + T_{\mu\nu}.
\end{equation}
Using junction condition \cite{Israel:1966rt}, one obtains the effective 4D Einstein equation on
a brane \cite{Shiromizu:1999wj}:
\begin{equation} \label{eq-induced}
  {}^{(4)} G_{\mu\nu} = -\Lambda_4 g_{\mu\nu} + \kappa^2 T_{\mu\nu} +  \kappa_5^4 \Pi_{\mu\nu} - {\cal E}_{\mu\nu}.
\end{equation}
In this equation, the quantities $\kappa$ and $\Lambda_4$, which are 4D gravitational coupling
constant and 4D cosmological constant, respectively, are given by relations:
\begin{equation}
 \Lambda_4 = \frac{1}{2} \left(\Lambda_5 + \frac{\kappa_5^4}{6} \sigma^2 \right), \;\;\;\;
 \kappa^2 = \kappa_5^2 \frac{\sigma}{6}.
\end{equation}
In AdS bulk, $\Lambda_5 < 0$. In addition, we will use the RS fine tuning condition:
\begin{equation} \label{ft}
 \Lambda_5 = - \frac{\kappa_5^4}{6} \sigma^2 ,
\end{equation}
which is necessary for static solutions to exist in RS2 model.
The 4D gravitational constant becomes
\begin{equation}
 \kappa^2 = \frac{\sigma \kappa_5^4}{6} = - \frac{\Lambda_5}{\sigma}.
\end{equation}
At last, the bulk Einstein equations, $G_{AB} = -\Lambda_5 g_{AB}$, give the relation between
the 5D cosmological constant and the AdS curvature radius, $\Lambda_5=-6/\ell^2$.

Further, the tensor $\Pi_{\mu\nu}$ in Eq. (\ref{eq-induced}) is given by the
expression
\begin{equation}
\label{Smunu}
  {\Pi}_{\mu\nu}= -{{1\over 4}} T_{\mu\alpha} T_{\nu}^{\alpha}
  +{{1\over12}}T_{\alpha}^{\alpha}T_{\mu\nu} + {{1\over24}}g_{\mu\nu}
  \left[3 T_{\alpha\beta} T^{\alpha\beta}-(T_\alpha^\alpha)^2 \right],
\end{equation}
and ${\cal E}_{\mu\nu}$ is the limiting value on the brane of the electric part of
the bulk Weyl's tensor. The latter term is, in the effective Einstein's equations, an
external source, with an energy-momentum tensor $T_{\mu\nu}^{\cal E}$, defined as
\begin{equation}
 T_{\mu\nu}^{\cal E} = - \frac{1}{\kappa^2} {\cal E}_{\mu\nu}, \;\; {T^{\cal E}_{\mu} } ^ {\mu} = 0.
\end{equation}
In the case, which we consider in the present paper, this tensor is equal to zero because
the Weyl tensor, $C_{ABCD}$, vanishes for an AdS bulk.

The $\Pi_{\mu\nu}$-tensor term in the effective Einstein equations
(components of this tensor are quadratic in $\rho$) leads to the following
modification of the Friedmann equation ($\kappa^2=8\pi G$):
\begin{equation}
\label{FrEq}
H^2 = \frac { 8 \pi G}{3} \rho \left( 1+ \frac{\rho}{2 \sigma} \right).
\end{equation}
Deriving this formula, the fine tuning condition, Eq. (\ref{ft}), and equalities
$K=0$, $M=0$ in Eq. (\ref{hr}) are used.
The solution of Eq. (\ref{FrEq}) for a radiation-dominated
state ($p=\rho/3$) on the brane is (see, e.g., \cite{Sendouda:2006nu})
\begin{subequations}
\label{aHrho}
\begin{align}
a(t) = &\; a_{eq} \frac{t^{1/4} (t+t_c)^{1/4}}{t_{eq}^{1/2}} \;,\;\;\; \\
H(t) = &\; \frac{2t+t_c}{4t(t+t_c)} \;,\;\;\; \\
\rho(t) = &\; \frac{3}{32 \pi G t (t+t_c)},
\end{align}
\end{subequations}
where $t_c\equiv \ell/2$. The conservation equation
has the same form as in 4D case:
\begin{equation}
\dot \rho = - 3 H (\rho + p).
\end{equation}

As one can see, at late times (i.e., at low energy density), the well-known
relations of 4D cosmology are recovered. The time dependence of the horizon mass in RS model is
(we put $c=1$)
\begin{equation}
\label{Mht}
M_h(t) \equiv \frac{4 \pi \rho}{3 H^3} = \frac{8t^2(t+t_c)^2}{G (2 t +t_c)^3}.
\end{equation}

The transition between the so-called high energy (HE) and low energy (LE) regimes happens at
the ``critical'' epoch, at which $H \ell = 1$, and horizon mass at this
time (this happens at $t=\frac{\ell}{2\sqrt{2}}=t_c/\sqrt{2}$) is
\begin{eqnarray}
M_h(t_c/\sqrt{2}) \approx 5\times 10^{25}\; {\rm g} \left( \frac{\ell}{0.1 \;{\rm mm} } \right).
\end{eqnarray}
The critical value of the comoving wave number, $k_c$, which
corresponds to this critical epoch, can then be written
using the known relations between the horizon mass and the comoving
wave number (see, e.g., \cite{Bugaev:2009zh}) as
\begin{eqnarray}
k_c \approx k_{eq} \left( \frac{M_h(t_c/\sqrt{2})} {M_{eq}} \right)^{-1/2}
\left( \frac{g_{*c}}{g_{*eq}} \right)^{-1/12} \approx \;\;\;\;\;\;\;\; \nonumber \\ \approx
3\times 10^{10}\; {\rm Mpc^{-1}}
\left( \frac{\ell}{0.1 \;{\rm mm} } \right)^{-1/2} \left( \frac{g_{*c}}{100} \right)^{-1/12},
\end{eqnarray}
where $k_{eq}$, $M_{eq}$ and $g_{*eq}$ are wave number, horizon mass and effective number of
relativistic degrees of freedom corresponding to the time of matter-radiation equivalence and $g_{*c}$ is
the number of relativistic degrees of freedom corresponding to the critical epoch.

The main parameter of the model, $\ell$, can be constrained by Newton's law tests in
table-top experiments. The most recent results \cite{Kapner:2006si, Geraci:2008hb} give the
following limit, which is very important for cosmological implications of the model:
\begin{equation}
\label{ell-bound}
\ell \lesssim (0.015 - 0.044) \;{\rm mm} .
\end{equation}
The corresponding constraints from astronomical observations are
somewhat weaker (see, e.g., \cite{Johannsen:2008tm}).

\subsection{Scalar perturbations}

The case when $M=0$ in Eq. (\ref{hr}) corresponds to a pure AdS bulk spacetime. It is known that in this case
a study of cosmological perturbations in the bulk and the brane is greatly simplified.
It was shown in \cite{Mukohyama:2000ui, Mukohyama:2001yp, Kodama:2000fa} that a
solution of the perturbed 5D Einstein equations in a vacuum AdS bulk, having
only metric perturbations,
\begin{equation}
  {}^{(5)\!}\delta G^A_B = 0,
\end{equation}
can be reduced to a solution of the evolution equation for the ``master variable'' $\Omega$
(which depends only on coordinates in the 2-dimensional orbit space, i.e., on $\tau, z$)
whereas all gauge-invariant metric perturbations in the bulk are written in terms of this $\Omega$.

In Poincare coordinate system (used above, in Eq. (\ref{eq4})) the wave equation governing
the evolution of the master variable in the bulk (the master equation) is
\begin{equation}
\label{Omega-eq}
 -\frac{\partial^2\Omega}{\partial \tau^2} + \frac{\partial^2\Omega}{\partial z^2}
 + \frac{3}{z} \frac{\partial\Omega}{\partial z} + \left( \frac{1}{z^2} -
 k^2 \right) \Omega = 0.
\end{equation}
Here, and everywhere below, we work with Fourier transforms (with respect to the
$x^i$'s) of $\Omega$ and all the perturbation functions.

The important boundary condition for $\Omega$ can be obtained from Israel's junction
conditions \cite{Israel:1966rt}. These conditions take the simplest form in a GN
coordinate system, in which the bulk metric is
\begin{equation}
{}^{(5)} ds^2 = g_{\mu\nu} dx^\mu dx^\nu + dy^2.
\end{equation}
The perturbed 5D metric in this system is given, in generalized
5D longitudinal gauge, by the expression
\begin{equation}
\label{gAB}
g_{AB} = \left(\begin{array}{ccc}
-n^2(1+2\tilde A) & 0 & n \tilde A_y \\
0 & a^2\left[(1+2 {\cal \tilde R})\delta_{ij} \right]& 0 \\
n \tilde A_y & 0 &1+2 \tilde A_{yy}
\end{array}\right).
\end{equation}
All quantities in Eq. (\ref{gAB}) and, in particular, $n, a$, are functions of GN
coordinates $t, y$. On the brane one has $n_b=1$, $a_b=a(t, y=0)$.
The functions $a(y,t)$ and $n(y,t)$ are known from the solution of the
Einstein equations in GN coordinate system \cite{Binetruy:1999hy}. Scalar
quantities $\tilde A$, $\tilde A_y$, $\tilde A_{yy}$, $\tilde {\cal R}$ in Eq. (\ref{gAB})
are gauge invariants. The formulas relating the derivatives in two coordinate systems
are given by
\begin{subequations}
\begin{align}
 \frac{\partial}{\partial y}  = & \frac{1}{a} \left( - \ell \frac{\dot a}{a} \frac{\partial}{\partial \tau} +
 \sqrt{1+ \left(\frac{\dot a}{a}\right)^2 \ell^2} \; \frac{\partial}{\partial z} \right),\\
 \frac{\partial}{\partial t}  = & \frac{1}{a} \left(
 \sqrt{1+ \left(\frac{\dot a}{a}\right)^2 \ell^2} \; \frac{\partial}{\partial \tau} -
  \ell \frac{\dot a}{a} \frac{\partial}{\partial z} \right).
\end{align}
\end{subequations}

Using the expressions for the junction conditions \cite{Koyama:2004ap}, we neglect in
them the terms with anisotropic stress perturbation in the perturbed energy-momentum tensor for
matter on the brane and, correspondingly, all terms containing the brane bending scalar
$\xi(t, x^i)$ (describing the perturbed position of the brane) in the expression for
the perturbed extrinsic curvature tensor. In this approximation, one can introduce the
following notations:
\begin{equation}
\Phi = \tilde A_b, \; \Psi=- \tilde {\cal R}_b,
\end{equation}
having in mind that these gauge invariant perturbations of the bulk metric
coincide, on the brane, with lapse and curvature perturbations in the conventional
4D cosmological perturbation theory.

Junction conditions give the expressions for matter
perturbations ($\delta \rho$, $\delta q$, $\delta p$) on the brane through the linear
combinations of gauge invariants and their derivatives and, therefore, through
the master variable $\Omega$ and its derivatives.
Using these expressions, one can obtain, for $\Omega$, a boundary condition on
the brane expressed through the gauge-invariant quantity $\Delta$ (defined below
in Eq. (\ref{delta-descr})) \cite{Deffayet:2002fn}:
\begin{equation}
\label{boundary}
 \left[ \frac{\partial \Omega}{\partial y }  + \frac{1}{\ell} \left(1 +
 \frac{\rho}{\sigma} \right) \Omega + \frac{6\rho a^3}{\sigma
 k^2} \Delta \right]_{b} = 0.
\end{equation}

Considering the perturbed effective Einstein equations,
\begin{equation}
 \label{G4}
 {}^{(4)}\delta G_{\mu\nu} = \kappa^2 \delta T_{\mu\nu} + \kappa_5^4 \delta \Pi_{\mu\nu}
   - \delta {\cal E}_{\mu\nu},
\end{equation}
one can parameterize the perturbations of ${\cal E}_{\mu\nu}$ in the
form \cite{Langlois:2000ph, Maartens:2000fg}:
\begin{subequations}
\begin{align}
 \delta     \mathcal{E}_{0}{}^0 & = \kappa^2 \delta\rho_\mathcal{E} Y, \\
 \delta  \mathcal{E}_i{}^0 & = \kappa^2 k Y_i \delta q_\mathcal{E}, \\
 \delta \mathcal{E}_i{}^j & = -\kappa^2 (\tfrac{1}{3} \delta\rho_\mathcal{E}
  Y \delta_i{}^j + k^2 \delta\pi_\mathcal{E} Y_i{}^j),
\end{align}
\end{subequations}
\begin{equation}
Y = e^{i\mathbf{k} \mathbf{x}}, \quad Y_i = -\frac{1}{k}
 \partial_i Y, \quad Y_{ij} = \frac{1}{k^2} \partial_i \partial_j Y +
 \frac{1}{3} \delta_{ij} Y,
\end{equation}
treating the trace free tensor $\delta {\cal E}_{\mu\nu}$ as an additional fluid source term
in Eq. (\ref{G4}) with a radiation-like equation of state. This assumption is in full
analogy with a case of the tensor $T_{\mu\nu}$, where one has
\begin{subequations}
\begin{align}
  \delta     T_{0}{}^0 & = - \delta\rho Y, \\
  \delta  T_i{}^0 & = - k Y_i \delta q, \\
  \delta T_i{}^j & = \delta\rho Y \delta_i{}^j + k^2 \delta\pi Y_i{}^j.
\end{align}
\end{subequations}
The perturbations of the Weyl fluid, $(\delta\rho_\mathcal{E}, \delta q_\mathcal{E}, \delta\pi_\mathcal{E})$,
transfer effects of the bulk metric perturbations (effects of ``KK degrees of freedom'') to the brane.

The solution of the perturbed equations (\ref{G4}) is a generalization of the results
of standard 4D cosmological perturbation theory. The corresponding formulas are derived in
\cite{Cardoso:2007zh} (in the approximation $\delta\pi=0$). These formulas express gauge
invariants $\Phi, \Psi$ in terms of the gauge invariant matter perturbation variables
$\Delta$ (which is a density contrast in the comoving gauge) and $V$ (which is a peculiar velocity
in the longitudinal gauge), as in the 4D perturbation theory.
These invariants are given by the relations (in the longitudinal gauge):
\begin{equation}
\label{delta-descr}
\rho \Delta = \delta\rho - 3 H \delta q, \;\;\;\;\; a(\rho+p)V = -k \delta q.
\end{equation}
There are two differences from the 4D theory: the formulas include 
{\it i)} ${\cal O}(\rho/\sigma)$ corrections and {\it ii)} KK corrections, i.e., the
terms proportional to $\delta \rho_\mathcal{E}, \delta q_\mathcal{E}$ and $\delta\pi_\mathcal{E}$.
These latter terms can be expressed through the master variable~\cite{Deffayet:2002fn}:
\begin{subequations}
 \label{deltarhoeps}
\begin{align}
  \delta \rho _\mathcal{E} & = \left( \frac{k^4 \Omega}{3 \kappa^2 a^5} \right)_b, \\
  \delta q _\mathcal{E}  & = \left(\frac{k^2}{3 \kappa^2 a^3} \left[\dot\Omega -
  \frac{\dot a}{a}\Omega \right]\right)_b, \\
  \delta \pi_\mathcal{E} & = \frac{1}{6\kappa^2 a^3} \left( 3 \ddot\Omega - 3{{\dot a \over a}}
  \dot \Omega + {{k^2\over a^2}}\Omega - {{3\over 2}} \kappa_5^2 (p+\rho)\Omega' \right)_b,
\end{align}
\end{subequations}
where the prime denotes $\partial_y$ and the dot denotes $\partial_t$.

Using the results of \cite{Cardoso:2007zh} and Eqs. (\ref{deltarhoeps}) one can easily obtain the
ordinary differential equation for the gauge invariant $\Delta$. In the approximation
$c_s^2 = \dot p/ \dot \rho =1/3$, $w=p/\rho = 1/3$, one has
\begin{equation}
\label{Delta-eq}
\ddot \Delta + H \dot \Delta +
 \left[ \frac{1}{3} \left(\frac{k}{a}\right)^2
  - \frac{4 \rho }{\sigma \ell^2}  - \frac{18 \rho^2 }{\sigma^2 \ell^2}
   \right] \Delta =
   \frac{4 k^4  }{9 \ell a^5} \Omega_{b}.
\end{equation}
This equation contains the term which is proportional to $\Omega_b$, in the right-hand side.
Therefore, this equation is connected with Eqs. (\ref{Omega-eq}) and (\ref{boundary}).

Another important gauge invariant is the curvature perturbation on uniform density slices.
It is defined by the relation $\zeta=\psi-H\delta\rho/{\dot \rho}$, where $\psi$ is the curvature
perturbation. The relation between $\zeta$ and $\Delta$ also contains $\Omega_b$:
\begin{equation}
\label{zeta}
\zeta =  \left[ \frac{1}{4}
+ \frac{3 \rho a^2(3 \rho + 2 \sigma)}{4 k^2\ell^2\sigma^2} \right]
\Delta + \frac{3 Ha}{4 k^2} \frac{d\Delta}{d\eta}
+ \frac{k^2}{6\ell a^3} \Omega_{b}.
\end{equation}

\section{The high energy regime}
\label{Sec-f}

Studying, in the high energy regime of radiation dominated era
(when, in particular, $H\approx \frac{\rho}{\sigma\ell}$, $\partial_y \approx -\partial_t$),
the dependence of $\Delta$ and
$\Omega$ on time before horizon crossing, by power series methods, and taking into account
only the dominant growing mode, one can obtain (at leading order in $k\eta$) the
result \cite{Cardoso:2007zh}:
\begin{equation}
\label{as}
\Delta^{as} \approx {4\over3}(k\eta)^2, \;\; \Omega_b^{as} \approx 3 \ell a_*^3 k^{-2} (k\eta)^3.
\end{equation}
Here, $a_*$ is the scale factor at time of Hubble horizon crossing. In the high energy regime
one has
\begin{equation}
\eta = {1\over {3aH}}, \;\;\; a = a_* (3 k \eta)^{1/3}.
\end{equation}
The connection between $a_*$ and the comoving wave number is
\begin{equation}
\label{astar}
a_* = {k\over H_*} = a_c \cdot (\sqrt{2}-1)^{1/3}  \left( {k_c \over k} \right) ^{1/3},
\end{equation}
\begin{eqnarray}
a_c \equiv a(t_c/\sqrt{2}) \approx 1.25 \; \Omega_R^{1/4} \left( \frac{H_0 \ell}{c} \right)^{1/2}
\approx \qquad \qquad \nonumber \\ \qquad
\approx 10^{-16} \left( \frac{\ell}{0.1 \; {\rm mm}} \right)^{1/2}.
\end{eqnarray}

One can rewrite Eqs. (\ref{as}) in the form:
\begin{equation}
\label{as2}
\Delta^{as} \approx {4\over 27} \left({a\over a_*} \right)^6, \;\;\;
{1\over \ell} \Omega_b^{as} \approx {1\over 9} \left({a\over a_*} \right)^9 {a_*^3 \over k^2}.
\end{equation}
As one can see from Eqs. (\ref{as2}) and (\ref{astar}), the value of $\Delta^{as}$ at Hubble horizon
crossing is constant and the corresponding value of $\Omega_b^{as}$ depends only on $k$.

We are interested in a behavior of $\Delta$ and $\Omega_b$ in a relatively short time interval,
from $a=a_*$ up to $a\lesssim 3a_*$. 
As the results of the numerical calculations show (see Sec. \ref{sec-results}), 
just near $a\approx 3 a_*$ the $\Omega_b, \Delta$-values reach
maximum. At later times the oscillations begin, and amplitudes of these oscillations
are equal, approximately, to the maximum magnitudes of $\Omega_b, \Delta$ reached at the previous period of the
smooth behavior.

Our key assumption is the following: the growth of $\Omega_b$ amplitude in the interval
$(a_* \div 3a_*)$ can be described, in the limit of large $k$, $k \gg k_c$,
by a function which does not depend on the comoving wave number $k$.
Namely, one assumes that
\begin{equation}
\label{ff}
{1\over \ell} \Omega_b = {1\over 9} {a_*^3 \over k^2} \left({a\over a_*} \right)^9
f_\Omega \left( \frac{a}{a_*} \right),
\end{equation}
in the asymptotical limit of the high energy regime, $k \gg k_c$.
The function $f_\Omega$ decreases with a growth of $a/a_*$ and it is assumed that
$f_\Omega(1)=1$.

According to this assumption, the time evolution of $\Omega_b$, starting from
the horizon re-entry, is the same for all comoving wave numbers $k$, and the $k$-dependence of $\Omega_b$
enters only through initial conditions at $a=a_*$. It may be justified as follows.
A general solution of the master equation is given by the expression \cite{Koyama:2004cf}
\begin{equation}
\Omega_b = \frac{\ell^3}{z} \int dm S(m) Z_0(mz) e^{-i \omega \tau},
\end{equation}
where $Z_0$ is the linear combination of Hankel functions and $S(m)$ is the
arbitrary function, $\omega=\sqrt{m^2+k^2}$. The variable $m$ has a physical sense
of the KK mass. 
In the high energy regime, when $H\ell \gg 1$, the physical sizes of perturbations, at horizon 
re-entry, are smaller than $\ell$, $a_*/k=1/H_* \ll \ell$. Correspondingly, $k \ll a_*/\ell$.
At the same time, it is well known that in the high energy regime the contribution to $\Omega_b$
from the massive KK modes is, in general, significant (in contrast with the low energy case), i.e.,
the characteristic values of $m$ contributing to the integral for $\Omega_b$ can be much larger
than $a_*/\ell$. So, in the high energy regime, characteristic $m$-values are of the same order
as $k$-values, or even larger, and the $k$-dependence of $\Omega_b$ can be effectively masked
(if $\omega_{char}=\sqrt{m^2_{char} + k^2} \approx m_{char}$).

The assumption (\ref{ff}) is used below, for numerical calculations of $\Delta$ in the region $k \gg k_c$.
The results of these calculations show (see Sec. \ref{sec-results} for details) that
in this region of $k$, the ratios $\Delta(a)/\Delta(a_*)$ are the same for different $k$
(for $a_* < a \lesssim 3 a_*$). It follows from here that, in addition to (\ref{ff}), one can assume that
\begin{equation}
\label{ffDelta}
\Delta = {4\over 27} \left({a\over a_*} \right)^6 f_\Delta \left( \frac{a}{a_*} \right), \;\;\; f_\Delta(1) = 1,
\end{equation}
in the same interval of $a$, $a_* \div 3 a_*$.

Substituting now the expressions (\ref{ff}) and (\ref{ffDelta}) for $\Omega_b$ and $\Delta$ 
in the equations used for the numerical calculations, one can see that all of them 
become {\it independent on $k$} (in the high energy regime).

In the high energy regime, when $a\sim t^{1/4}$, the following useful relation holds:
\begin{equation}
H^2 a^2 = \left( \frac{a}{a_*} \right)^{-6} k^2.
\end{equation}
Using this relation and ansatzes (\ref{ff}) and (\ref{ffDelta}), one obtains, from Eq. (\ref{zeta}) for $\zeta$:
\begin{eqnarray}
\zeta = {1\over 27} \left({a\over a_*} \right)^6 f_\Delta \left( \frac{a}{a_*} \right) +
{1\over 3} f_\Delta \left( \frac{a}{a_*} \right) + \nonumber
\\
+ {1\over 9}\left[ 6 f_\Delta \left( \frac{a}{a_*} \right) +
 \frac{a}{a_*} f_\Delta' \left( \frac{a}{a_*} \right) \right] + \\ 
\nonumber + {1\over 54} \left({a\over a_*} \right)^6 f_\Omega \left( \frac{a}{a_*} \right).
\end{eqnarray}
At $a=a_*$, one has, as it must be, $\zeta \approx 1$, if the
condition 
\begin{equation}
\label{fINEQ1}
\frac{a}{a_*} f_\Delta'\left( \frac{a}{a_*} \right) \ll f_\Delta \left( \frac{a}{a_*} \right)
\end{equation}
holds. At $a/a_*=3$, i.e., near the maximum,
one has
\begin{equation}
\zeta_{max} = \zeta\left(\frac{a}{a_*}\approx3\right) = \frac{1}{27} 3^6
\left( f_\Delta(3) + {1\over2} f_\Omega(3) \right).
\end{equation}
If $f_\Delta(3)\approx f_\Omega(3) \sim 0.15$, one obtains that $\zeta_{max}\approx 6$.
This value is close to a value of the enhancement factor (see Sec. \ref{sec-results}).

The equation (\ref{Delta-eq}) for $\Delta$, after substituting of Eqs. (\ref{ff}, \ref{ffDelta}), becomes
\begin{eqnarray}
\label{ffD}
3 \frac{a}{a_*} f_\Delta'' + 30 f_\Delta' + \left( \frac{a}{a_*} \right)^5 f_\Delta =
\left( \frac{a}{a_*} \right)^5 f_\Omega.
\end{eqnarray}
Neglecting in Eq. (\ref{ffD}) the terms with derivatives, in accordance with Eq. (\ref{fINEQ1}),
one obtains the approximate result
\begin{eqnarray}
\label{fOfD}
f_\Delta\left( \frac{a}{a_*} \right) \approx f_\Omega\left( \frac{a}{a_*} \right).
\end{eqnarray}

Analogously, from the equation (\ref{boundary}) for the boundary condition one obtains
\begin{eqnarray}
\label{ffder}
\frac{a}{a_*} f_\Omega' - 8 f_\Delta + 8 f_\Omega = 0.
\end{eqnarray}
For consistency with Eq. (\ref{fOfD}), the function $f_\Omega(a/a_*)$ must obey the inequality
\begin{eqnarray}
\frac{a}{a_*} f'_\Omega\left( \frac{a}{a_*}\right) \ll f_\Omega\left( \frac{a}{a_*}\right).
\end{eqnarray}
This condition is consistent with the similar condition (\ref{fINEQ1}) and with (\ref{fOfD}).

In conclusion, we showed in this section, that the assumptions (\ref{ff}) and (\ref{ffDelta})
lead to the independence of the ratio $\Delta(a)/\Delta(a_*)$ on $k$ in the interval $a_* \div 3 a_*$,
in the high energy regime. Estimates show that for the mode with $k=10 k_c$ the critical epoch corresponds
to the moment of time when $a=3 a_*$. Therefore, beginning from $k \approx 10 k_c$, the interval
$(a_* \div 3 a_*)$ is entirely in the high-energy regime. Correspondingly, the asymptotical region in
which $\Delta(a)/\Delta(a_*)$ is independent on $k$ begins from $k$'s which are larger than $10 k_c$ (say, from 
$k \sim 30 k_c$).

\begin{figure}
\center %
\includegraphics[width= 8 cm, trim = 0 5 0 0 ]{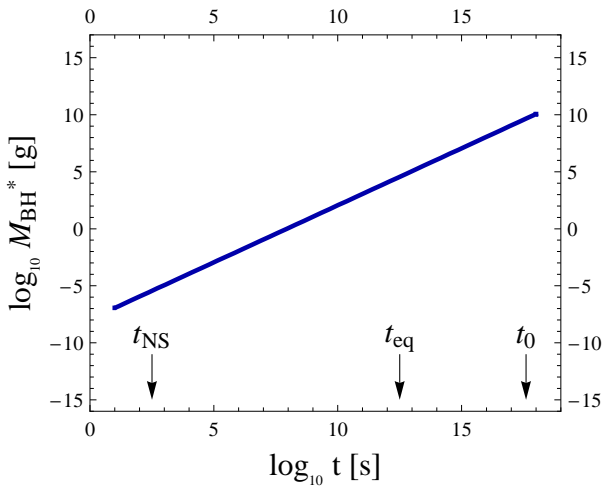}
\caption{ \label{fig-Mh-S-t} Black hole mass $M_{BH}^*$ versus the moment of time at which
it evaporates, assuming the case of RS cosmology with $\ell = 0.1 \;{\rm mm}$.
Labels ``$t_{NS}$'', ``$t_{eq}$'' and ``$t_0$'' show, correspondingly, the nucleosynthesis,
matter-radiation equivalence and present epochs.} %
\end{figure}

\section{Characteristics and evolution of 5D black holes}
\label{sec-5D}

The formation and evolution of PBHs in RS2 cosmology had been investigated in \cite{Guedens:2002km,
Guedens:2002sd, Majumdar:2002mra, Clancy:2003zd, Sendouda:2003dc, Majumdar:2005ba, Sendouda:2006nu}.

\begin{figure}[!b]
\includegraphics[width=8 cm, trim = 0 0 0 0 ]{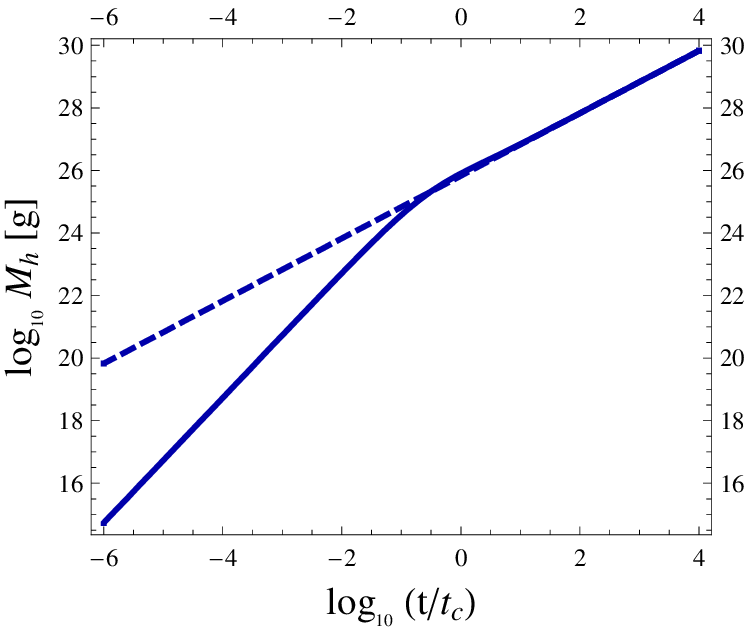} \;\;\;\;\; \\
\includegraphics[width=8 cm, trim = 0 0 0 0 ]{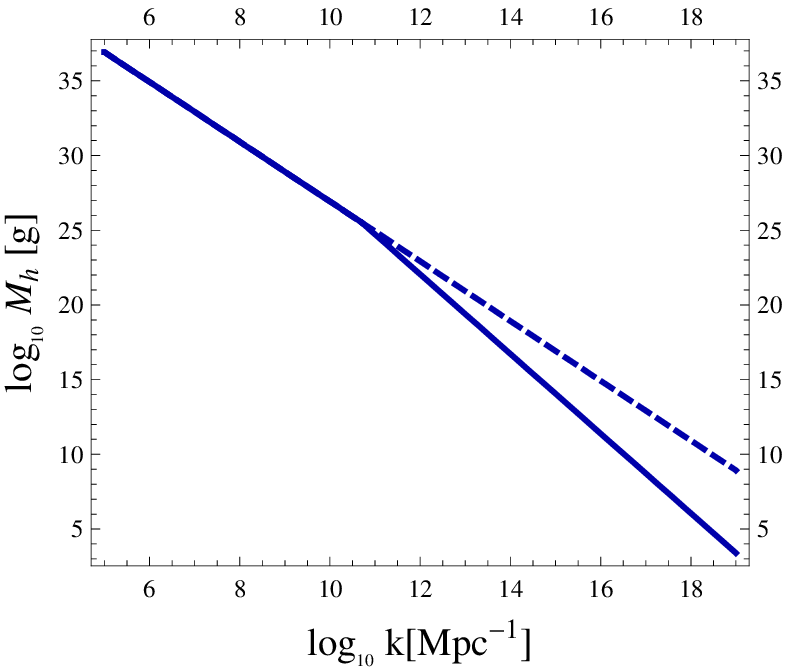}
\center %
\caption{ \label{fig-Mh} The dependence of horizon mass on cosmic time $t$
{\it (upper panel)} and comoving wave number $k$ {\it (lower panel)}. Solid curves show the case of RS cosmology
assuming that $\ell = 0.1 \;{\rm mm}$ while dashed curves are for the case of standard 4D-cosmology.} %
\end{figure}

Supposing that the braneworld PBHs localized on the brane are represented
by the 5D Schwarzschild solution \cite{Tangherlini:1963bw, Myers:1986un} for the
metric, one obtains, for the induced 4D metric on the brane, the expression
\begin{equation}
ds^2 = - \left[ 1-\left( \frac{r_s}{r} \right)^2 \right] dt^2 +
 \left[ 1-\left( \frac{r_s}{r} \right)^2 \right]^{-1} dr^2 + r^2 d\Omega^2,
\end{equation}
which does not coincide with the 4D Schwarzschild metric.
Correspondingly, the relation between PBH mass $M_{BH}$ and radius,
\begin{equation}
\label{rs5}
r_s = \sqrt{\frac{8}{3\pi}} \left( \frac{\ell}{\ell_4} \right)^{1/2}
 \left( \frac{M_{BH}}{M_4} \right)^{1/2} \ell_4,
\end{equation}
is different from the analogous relation in the 4D case (throughout this section,
we will use, following \cite{Guedens:2002km}, the convenient notations,\
in which $M_4$ is the Planck mass, $\ell_4=M_4^{-1}$ is the Planck length,
$t_4=\ell_4$ is the Planck time).

It follows from Eq. (\ref{rs5}) that if PBH's radius at its formation
is smaller than AdS radius $\ell$, the following inequality for PBH's mass
holds:
\begin{equation}
\frac{M_{BH}}{M_4} < \frac{\ell}{\ell_4}. \label{ineq}
\end{equation}
We assume, as usual, that PBHs form with masses equal, approximately, to the horizon mass
$M_h$ at the time of formation,
\begin{equation}
M_{BH}\approx M_h. \label{MBHMh}
\end{equation}
Using the expression for $M_h$ (Eq. (\ref{Mht})) and the relation $t_c=\ell/2$, one can see
that the equality (\ref{MBHMh}) is consistent with the inequality (\ref{ineq}) only if
$t<t_c$, i.e., in the high energy regime. It means that PBHs which form in the high energy regime
are 5D black holes.

A rate of a loss of the PBH's mass, due to the 5D-evaporation, is proportional to
$r_s^{-2}$ for the evaporation in the brane as well as in the bulk. So, one has
$dM_{BH}/dt \sim M_{BH}^{-1}$. Resulting lifetime of the black hole, $t_{evap}$, is proportional
to $M_{BH}^2$ rather than $M_{BH}^3$ as in the 4D case,
\begin{equation}
\label{tevap}
\frac{t_{evap}}{t_4} \sim \frac{\ell}{\ell_4} \left(\frac{M_{BH}(t_c, t_{evap})}{M_4} \right)^2.
\end{equation}
In this formula $t_c$ is the time of the onset of evaporation, $M_{BH}(t_c, t_{evap})$ is the
PBH mass at $t=t_c$ which evaporates at $t=t_{evap}$ (one assumes that $t_c \ll t_{evap}$).
Here we assume, following \cite{Clancy:2003zd}, that in the relatively short period
of time from the PBH formation, $t_i$, up to the end of the high energy regime, $t_c$,
black hole does not evaporate, but increases its mass due to the accretion. The increase
of mass due to the accretion is determined by the equation \cite{Majumdar:2002mra, Guedens:2002sd}
$dM/dt \sim q M/t$ ($q$ is the (unknown) parameter of an efficiency of the accretion, $0<q<1$).
It is assumed, for simplicity, in a derivation of Eq. (\ref{tevap}) that at $t=t_c$
the accretion process ends completely, giving place for the pure evaporation.

One can check that if AdS radius is too small, the condition for 5-dimensionality of
PBHs, $r_s < \ell$, can not be satisfied. Comparing the mass-lifetime relation (\ref{tevap})
with the expression for the radius (\ref{rs5}), one can determine, for a given
lifetime, the minimal possible value of $\ell$, given by the relation
\begin{equation}
\ell_{min} \sim \left( \frac{t_{evap}}{t_4} \right)^{1/3} \ell_4.
\end{equation}
If, e.g., PBH evaporates today, $t_{evap}\approx t_0 \sim 10^{17}\;$s, one has
$\ell_{min} = 10^{20} \ell_4$.

The BH mass at the onset of the evaporation ($t=t_c$), if the age of the black hole is equal
to the age of the Universe, is
\begin{eqnarray}
M_{BH}(t_c,t_0) \equiv M_{BH}^*(t_0) \approx 5 \times 10^9
\left( \frac{\ell}{0.1\; {\rm mm}} \right)^{-1/2} {\rm g}, \nonumber \\
 \ell > 10^{20} \ell_4. \quad
\end{eqnarray}
and, in general, for PBH evaporating at time $t$,
\begin{eqnarray}
M_{BH}^*(t) \approx 5 \times 10^9
\left( \frac{\ell}{0.1\; {\rm mm}} \right)^{-1/2} \left( \frac{t}{t_0} \right)^{1/2} {\rm g},
\nonumber \\
\ell > \left( \frac{t}{t_4} \right)^{1/3} \ell_4. \quad
\end{eqnarray}
The dependence of $M_{BH}^*$ on $t$ for $\ell = 0.1\;$mm is shown in Fig. \ref{fig-Mh-S-t}.

\begin{figure}
\center %
\includegraphics[width=8 cm]{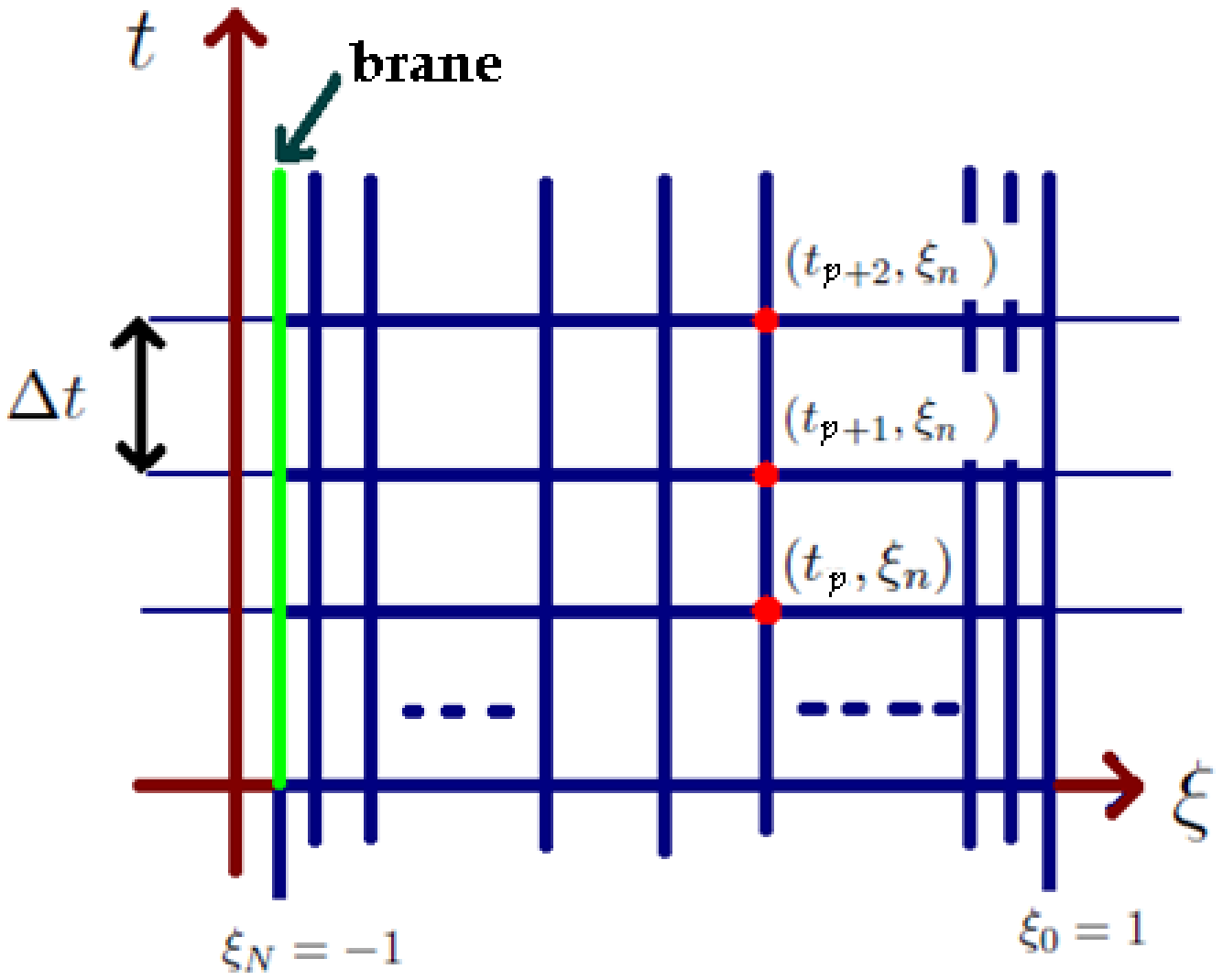} \;\;\; \\
$\;$ \\
\includegraphics[width=8 cm]{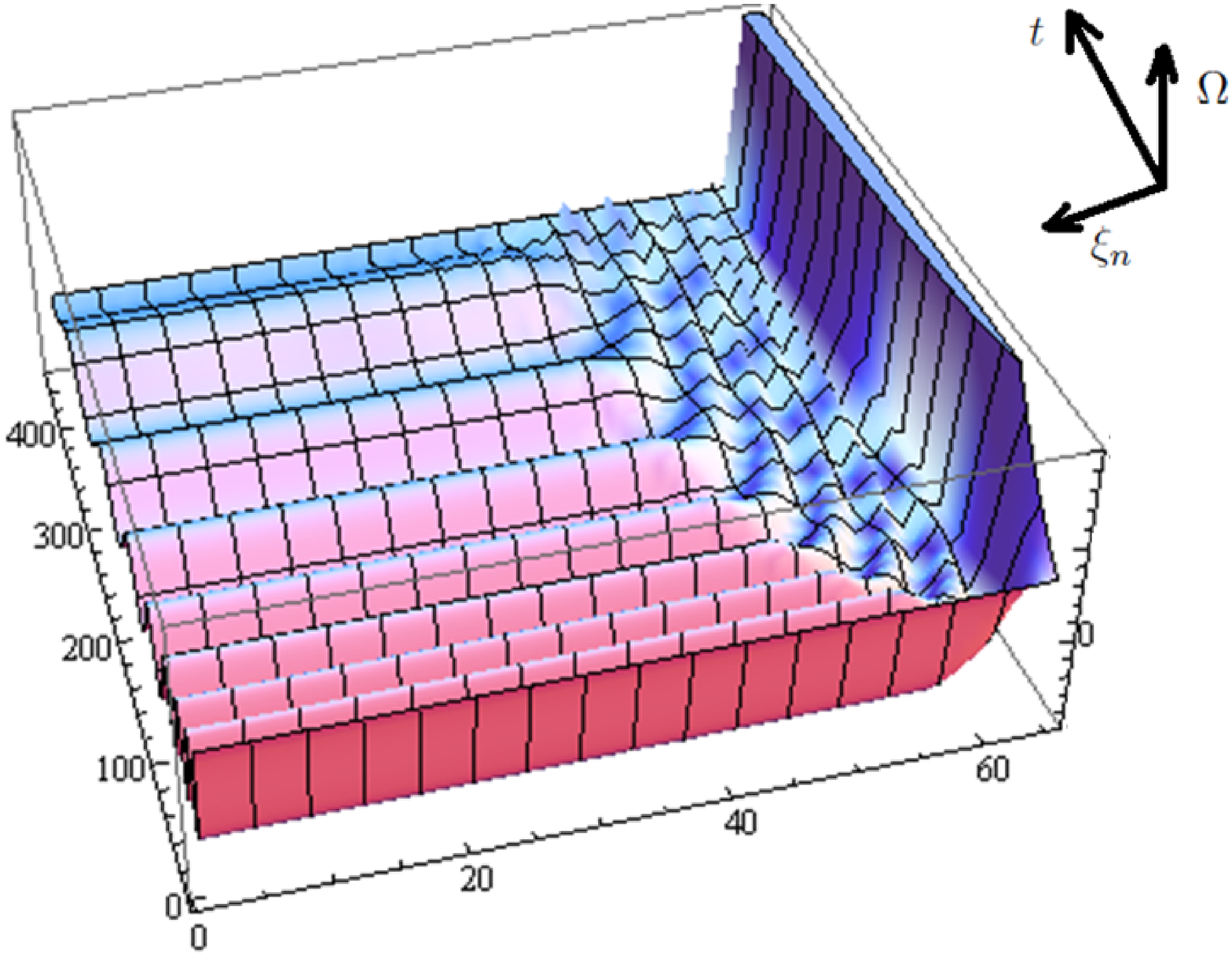}
\caption{ \label{fig-grid} {\it (upper panel)} The computational grid on which we solve the system of equations
(\ref{Delta-eq}) and (\ref{Omega-eq-2}) with boundary condition (\ref{boundary}).
The physical brane is at $\xi=\xi_N=-1$ while the regulatory one is at $\xi=\xi_0=1$.
{\it (lower panel)} An example of computational result for master variable $\Omega$ (in this case,
we took $k=30 k_c$ and $N=64$, so brane is at $\xi=\xi_{64}=-1$). The grid over $n$ is shown
to be homogenous, however, note, that the actual $\xi_n$-grid is inhomogeneous due to Eq. (\ref{Ga-Lo}).}
\end{figure}

One should note, in a conclusion of this section, that if the accretion efficiency is not small,
the initial masses of the PBHs are smaller than its masses at the onset of the evaporation.
And, even without any accretion, initial masses of the 5D PBHs are, for the same values of
total lifetime, much smaller than initial masses in the standard cosmology. It means
that in the 5D case the known astrophysical constraints on the PBH abundance correspond to
primordial perturbations on smaller scales.

The dependence of horizon mass on time and $k$ in brane cosmology is shown in Fig. \ref{fig-Mh}.
It is seen that PBHs with mass less than $\sim 10^{25}\;$g are produced in the
high energy regime. The corresponding comoving wave numbers are larger than
$10^{11}\;$Mpc$^{-1}$. If $\ell \approx 0.1\;$mm then PBHs evaporating today
correspond to $k\gtrsim 10^{17}\;$Mpc$^{-1} \sim 10^6 k_c$.

\section{Numerical scheme}
\label{sec-scheme}

It follows from Sec. \ref{sec-5D} that in brane cosmology PBHs having, at formation, relatively
small masses ($\lesssim (10^9-10^{10})\;$g) and, in particular, evaporating near today, had been produced
long before the critical epoch, $t_{form} \ll t_c$. The corresponding comoving sizes of
perturbed regions are also small, $k^{-1} \sim (10^{-16} - 10^{-17})\;$Mpc. Therefore, it is
practically important to determine the enhancement factors for rather large values of $k$,
$k\gtrsim (10^6-10^7) k_c$.
The straightforward numerical calculation of these factors for such large $k$ are quite difficult
and unreliable, but, luckily, the numerical calculations for moderately large $k$, $k\sim (10-30)k_c$,
show the flattening and the possible saturation of the dependence of the enhancement factor $\cal Q$
on $k$. We argue now that, really, ${\cal Q}(k)$ does not depend on $k$ in the limit of very large $k$,
$k \gg k_c$.

For the numerical solution of the system of equations (\ref{Omega-eq}) and (\ref{Delta-eq})
with boundary condition (\ref{boundary}), a pseudo-spectral calculation method was employed.
Such methods are often used in the tasks of hydrodynamics and a detailed description
can be found, e.g., in \cite{Canuto}.

To be able to perform a spectral transformation over the set of Chebyshev polynomials we
do a following change in the variables:
\begin{equation}
\Omega(\tau, z) \to \Omega(t, \xi),
\end{equation}
\begin{equation}
\xi = \frac{2 z - (z_{reg}+z_b(t))}{z_{reg}-z_b(t)}, \;\;\; -1 \le \xi \le 1,
\end{equation}
where $t$ has the meaning of cosmic time on the brane and is related to $\tau$ by
\begin{equation}
\frac{d\tau}{dt} = \frac{\sqrt{1+\ell^2 H(t)^2}}{a(t)} ,
\end{equation}
while $z_{reg}$ is the position of the regulatory boundary (artificial cutoff that is introduced
to make a computational domain finite, see, e.g., \cite{Hiramatsu:2006bd, Hiramatsu:2006cv}).

The equation for $\Omega$ (\ref{Omega-eq}) is rewritten in new variables as
\begin{equation}
\label{Omega-eq-2}
\frac{\partial^2 \Omega}{\partial t^2}  +
  K_{t\xi} \frac{\partial^2 \Omega}{\partial t \partial \xi} +
  K_{\xi\xi} \frac{\partial^2 \Omega}{ \partial \xi^2}
  + K_t \frac{\partial \Omega}{\partial t} +
 K_\xi \frac{\partial \Omega}{\partial \xi}+ K \Omega = 0,
\end{equation}
where
\begin{subequations}
\begin{align}
 K_{t\xi}(t,\xi) & = 2 \frac{\partial \xi}{\partial \tau} \left( \frac{dt}{d\tau} \right)^{-1}, \\
 K_{\xi\xi}(t,\xi) & = \left[ \left( \frac{\partial\xi}{\partial \tau}\right)^2 -
  \left(\frac{2}{z_{reg}-z_b(t)} \right)^2 \right]  \left( \frac{dt}{d\tau} \right)^{-2}, \\
 K_t(t,\xi) & = 2 \frac{d^2 t}{d \tau^2} \left( \frac{dt}{d\tau} \right)^{-2}, \\
 K_\xi(t,\xi) & = \left( 2 \frac{\partial^2\xi}{\partial \tau^2} -
     \frac{6}{z(t,\xi)(z_{reg}-z_b(t))} \right) \left( \frac{dt}{d\tau} \right)^{-2}, \\
 K(t,\xi) & = - \left( \frac{1}{z(t,\xi)^2} - k^2 \right)   \left( \frac{dt}{d\tau} \right)^{-2}.
\end{align}
\end{subequations}
Further, a new variable $\chi$ related to time derivative of $\Omega$ is introduced \cite{Hiramatsu:2006bd}
to reduce the task to two first-order equations:
\begin{equation}
\label{PDE1}
\frac{\partial \Omega}{\partial t} = \chi - K_{t\xi} \frac{\partial \Omega}{\partial \xi}
 \equiv F(\chi, \Omega'_\xi; \; t, \xi),
\end{equation}
\begin{eqnarray}
\label{PDE2}
\frac{\partial \chi}{\partial t} = -K_{\xi\xi} \frac{\partial^2 \Omega}{ \partial \xi^2} +
\left( \frac{\partial K_{t\xi}}{ \partial t} - K_\xi \right) \frac{\partial \Omega}{ \partial \xi} -
 \nonumber \\ -
K_t \left( \chi - K_{t\xi} \frac{\partial \Omega}{ \partial \xi} \right) - K \Omega
\equiv  \nonumber \\ \equiv
G(\chi, \Omega, \Omega'_\xi, \Omega''_{\xi\xi}; \; t, \xi).
\end{eqnarray}

To solve this system using difference method,
the transformation of all quantities over set of Chebyshev polynomials is done for the $\xi$ (and $\chi$) - axis.
This is done at each time step so that system of partial
differential equations (\ref{PDE1}, \ref{PDE2}) reduces to the system of ordinary
differential equations.

Thus, at each point $(t_p,\xi_n)$, the following quantities are known:
$$ \chi, \Omega, \Omega'_\xi, \Omega''_{\xi\xi}, F, G, $$
and also known are Chebyshev transforms
$$ \tilde \chi_n, \tilde \Omega_n, (\tilde \Omega'_\xi)_n, (\tilde \Omega''_{\xi\xi})_n, \tilde F_n, \tilde G_n. $$
The grid (see Fig. \ref{fig-grid} for illustration) based on Gauss-Lobatto points is used here,
\begin{equation}
\label{Ga-Lo}
\xi_n = \cos\left( \frac{\pi n}{N} \right), \;\;\;\;\; n=0,1,...,N,
\end{equation}
because it allows to perform fast Fourier transforms (FFTs) between the set of values of any
variable (e.g., $\Omega$) in Gauss-Lobatto points and its Chebyshev components $\tilde \Omega_n$.
Chebyshev transforms of derivatives, such as $(\tilde \Omega'_\xi)_n$ and $(\tilde \Omega''_{\xi\xi})_n$,
are also easily obtained using recurrence relations from the Chebyshev components of the
function (see \cite{Canuto} for details).

\begin{figure}
\center %
\includegraphics[width=8 cm]{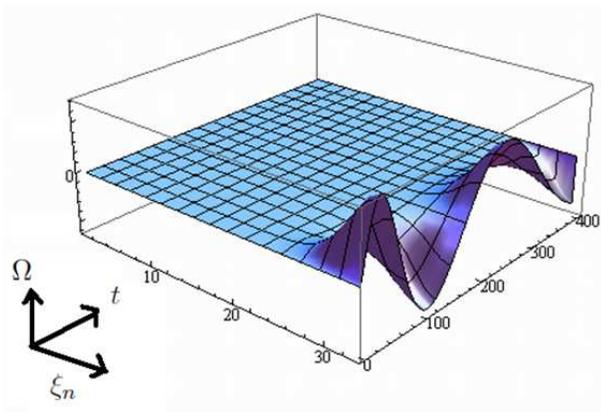} \;\;\;\; \\
$\;$ \\
\includegraphics[width=7.2 cm, trim = 0 5 0 0 ]{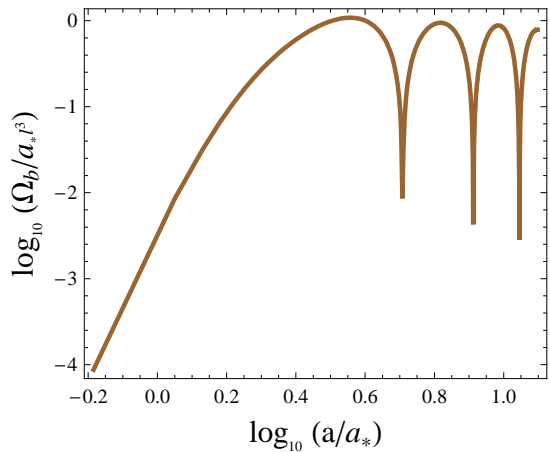}
\caption{ \label{fig-Om-3kc} The result of the numerical calculation of master
variable $\Omega$ for $k=3 k_c$. {\it Upper panel:} the
illustration of $\Omega(\xi_n,t)$ (here, $N=32$, units over time and $\Omega$-axis are arbitrary);
{\it lower panel:} the value of $\Omega$ on the brane, $\Omega_b$. The normalization is
given by $\zeta=1$ for $a\ll a_*$.} %
\end{figure}

\begin{figure}[!b]
\includegraphics[width=7 cm, trim = 0 0 0 0 ]{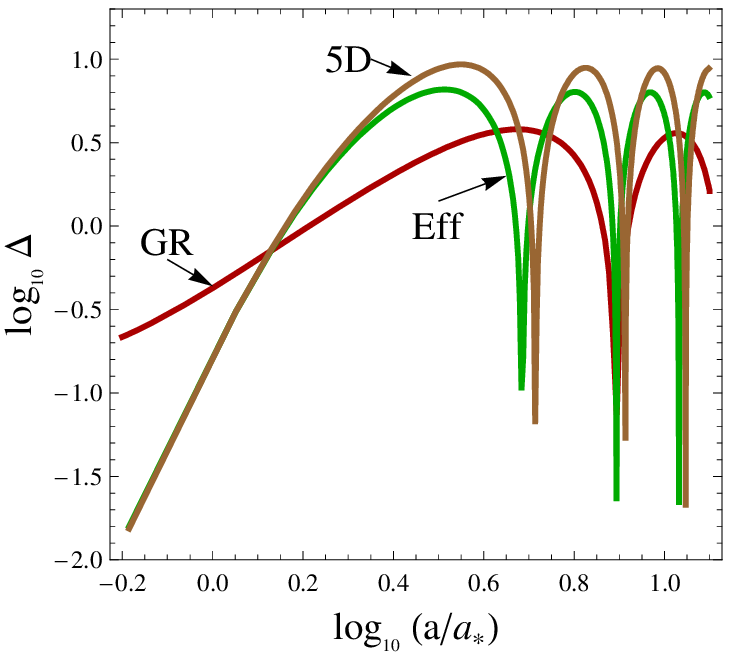} \\
\includegraphics[width=7.3 cm] {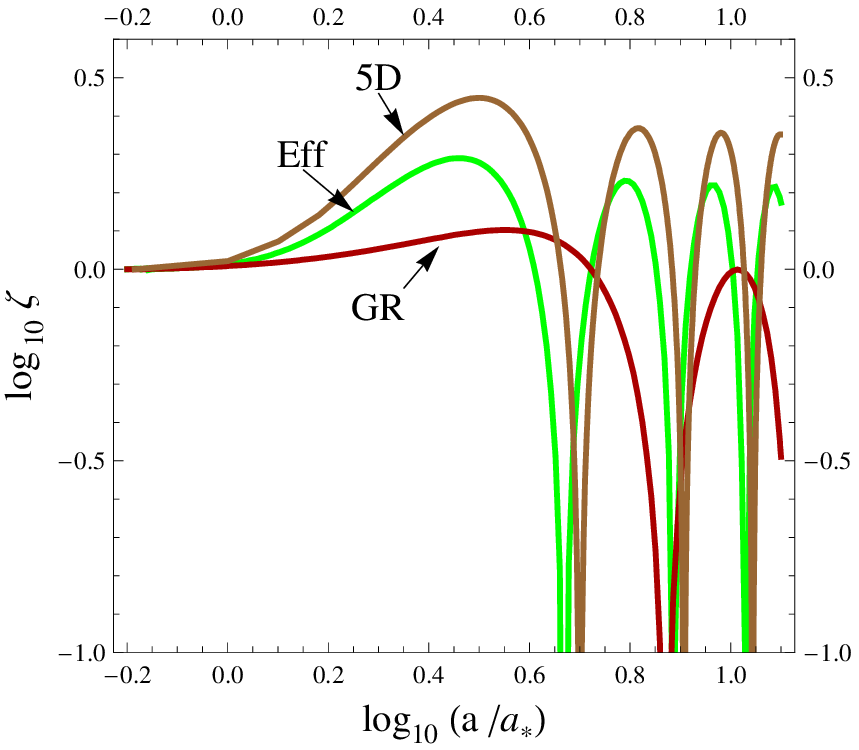}
\center %
\caption{ \label{fig-3kc-delt} The calculation of the density contrast and curvature
perturbation on the brane for $k=3 k_c$ for the cases of
full 5-dimensional calculation (curves labeled ``5D''), 
effective approach (approximation of $\Omega_b=0$, curves labeled ``Eff'') and
General Relativity, i.e. standard cosmology (``GR'').
Note that the value of $a_*$ is the same for ``5D'' and ``Eff'' cases, but is different for ``GR''
because of the different evolution of the background quantities.
{\it Upper panel:} Comoving density contrast $\Delta$ as a function of the scale factor,
normalized to $\zeta=1$ in super-horizon regime.
{\it Lower panel:} Curvature perturbation $\zeta$ calculated using the same
three approaches. } %
\end{figure}

Equations that are actually solved on each time step are:
\begin{equation}
\frac{d\tilde\Omega_n}{dt} = \tilde F_n (t); \;\;\; \frac{d\tilde\chi_n}{dt} = \tilde G_n (t).
\end{equation}
For points on the brane, $\Delta(t_p)$ is also evaluated at each step and Eq. (\ref{Delta-eq}) is solved
using a finite difference method (in our case, a 4-th order Adams-Bashforth-Moulton scheme).

The boundary conditions are imposed on the values of the highest two components of the
master variable, $\tilde\Omega_N$ and $\tilde\Omega_{N-1}$.
This is done by demanding the following:
\begin{subequations}
\begin{align}
\Omega'_\xi (-1) &= \sum_{n=0}^{N} \tilde \Omega_n T_n'(-1), \\
\Omega'_\xi (1) &= \sum_{n=0}^{N} \tilde \Omega_n T_n'(1),
\end{align}
\end{subequations}
where $T_n(\xi)$ is the $n$-th order Chebyshev polynomial. The value of $\Omega'_\xi (1)$ is
assumed to be zero (condition on the regulatory brane) while the value on the physical
brane, $\Omega'_\xi (-1)$, is related to other quantities by the boundary condition (\ref{boundary})
which can be expanded as
\begin{equation}
\left( \frac{\partial\Omega}{\partial \xi} \right)_{\xi=-1} =
\left(
 \frac
  {\frac{H\ell \chi}{\sqrt{1+H^2\ell^2}} - \frac{1}{\ell}\left(1+\frac{\rho}{\sigma}\right) \Omega -
    \frac{6\rho a^3 \Delta}{\sigma k^2}}
  {\frac{2 \sqrt{1+H^2\ell^2} }{a(z_{reg} - z_b(t)) }- \frac{H\ell}{a} \frac{\partial\xi}{\partial \tau} +
   \frac{ K_{t\xi}H \ell} {\sqrt {1+H^2\ell^2} } }
\right) _ {\xi = -1 }.
\end{equation}
Following the approach of \cite{Hiramatsu:2006bd}, we also use the following additional condition:
$\tilde\chi_N=\tilde\chi_{N-1}=0$.

\section{Results and conclusions}
\label{sec-results}

\begin{figure}
\center %
\includegraphics[width=8 cm]{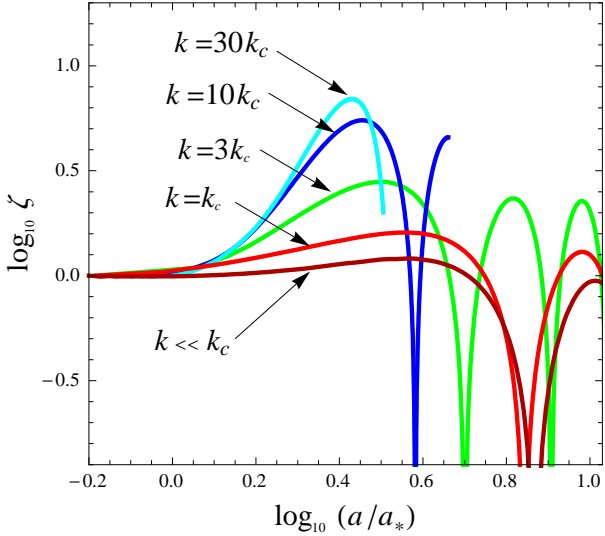}
\caption{ \label{fig-zeta-k} The result of the numerical calculation of $\zeta$ for
different values of $k$.} %
\end{figure}

\begin{figure}
\center %
\includegraphics[width=8 cm]{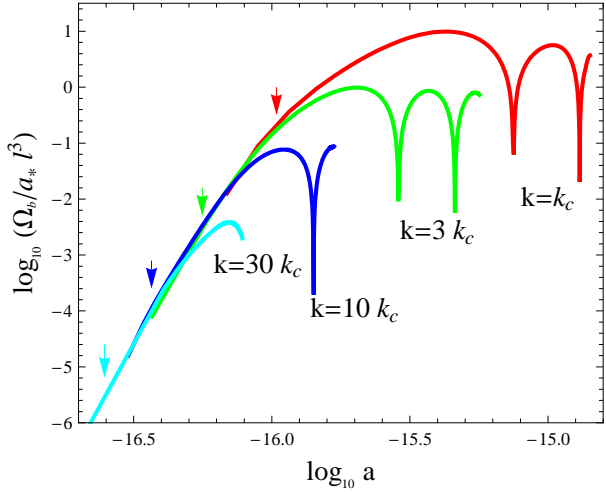}
\caption{ \label{fig-Omega} The result of the numerical calculation of $\Omega_b$ for
different values of $k$, normalized to $\zeta_k=1$ in the super-horizon regime.
Arrows show the horizon crossing time ($k=aH$) for each mode.} %
\end{figure}

\begin{figure}[!b]
\center %
\includegraphics[width=7.5 cm]{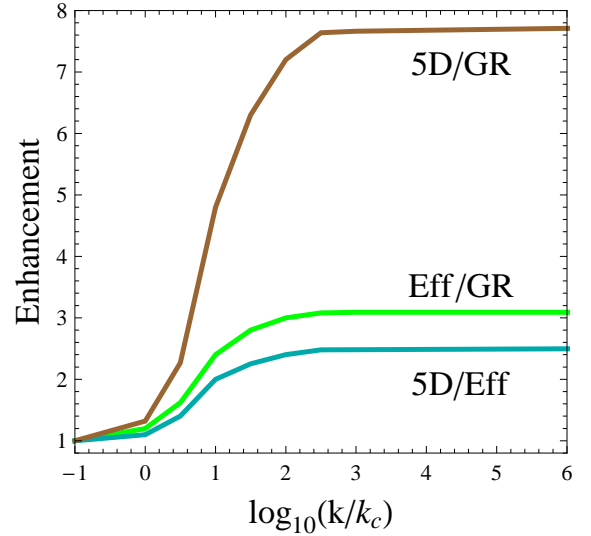}
\caption{ \label{fig-enh} Enhancement factors that show the degree of increasing of the
perturbation amplitude after horizon entry. From bottom to top, curves show the enhancement
of the amplitude of 5-dimensional calculation compared to the effective one, effective
theory compared to General Relativity result and 5-dimensional calculation compared to
General Relativity result.} %
\end{figure}

The main results of the numerical calculations are shown in Figs. \ref{fig-Om-3kc} - \ref{fig-enh}.
Figs. \ref{fig-Om-3kc}, \ref{fig-3kc-delt} show, for a given value of $k$, $k=3k_c$, the evolution,
near horizon crossing, of three variables: $\Omega_b$ (Fig. \ref{fig-Om-3kc}), $\Delta$ (Fig. 
\ref{fig-3kc-delt}, upper panel) and $\zeta$ (Fig. \ref{fig-3kc-delt}, lower panel). It is clearly 
seen that all three variables rise with $a$, up to $a\sim 3 a_*$. It is also seen from Fig. \ref{fig-3kc-delt}
that the corresponding rise of $\Delta$ and $\zeta$ in GR is weaker, and, as a result, we have an
enhancement. It follows also, from Fig. \ref{fig-3kc-delt}, that the enhancement is not zero
in the approximation $\Omega_b=0$, when there are no KK corrections in the equations of the 5D perturbation
theory.

In Figs. \ref{fig-zeta-k}, \ref{fig-Omega} it is shown how the evolution curves for $\zeta$ and
$\Omega_b$ change with an increase of $k$. It is seen, from Fig. \ref{fig-zeta-k}, that 
the enhancement grows with $k$, but there is a clear tendency of a slowdown of this growth
at $k > 10 k_c$.

Following \cite{Cardoso:2007zh} we define the factors that show the degree of 
enhancement of the perturbation amplitudes:
\begin{equation}
{\cal Q}_{eff} = \frac {\Delta_{eff}} {\Delta_{GR}} \;, \;
{\cal Q}_{\mathcal{E}} = \frac {\Delta_{5D}} {\Delta_{eff}} \;, \;
{\cal Q}_{5D} = \frac {\Delta_{5D}} {\Delta_{GR}}  = {\cal Q}_{eff} {\cal Q}_{\mathcal{E}}.
\end{equation}
In a case of the effective theory ($\Omega_b=0$), the enhancement reaches an asymptotic value,
${\cal Q}_{eff} \approx 3$, at $k\sim 100 k_c$. However, the direct calculation of ${\cal Q}_{5D}$ (or,
equivalently, ${\cal Q}_{\mathcal{E}}$) for very large wave numbers $k\gg k_c$ is not
easy, due to a quite complicate behavior of $\Omega$ in the bulk (see Fig. \ref{fig-grid} for
an illustration: the larger value of $k$, the more frequent are the oscillations in the bulk).
Due to limitations of computing resources, we have been able to make direct calculations in 5D case
only for a limited range of $k\lesssim 30 k_c$.

In order to study the PBH production for masses $M_{BH}^*(t_0) \sim 10^9\;$g
(such PBHs, as we have seen in Sec. \ref{sec-5D}, evaporate near
today, if the value of $\ell$ is close to its upper bound (\ref{ell-bound})),
we need information about cosmological perturbations for $k \gtrsim 10^6 k_c$.
To perform calculations for such large wave numbers, we have used an approximate
approach according to which $\Omega_b(a/a_*)$ has the same form 
(and is given by Eq. (\ref{ff})) for all large wave numbers
(see the discussion in Sec. \ref{Sec-f}).
In these calculations we have used, for the required function $f_\Omega(a/a_*)$, 
the corresponding function obtained from the direct numerical calculation for $k=30 k_c$.
Using this approach, we have calculated the enhancement factors for large values of $k$
($k \gtrsim 30 k_c$). The results of the calculation are shown in Fig. \ref{fig-enh}.

In summary, we stressed in this paper that, in RS2 brane cosmology, the PBHs of relatively small mass
(the concentration of which in space can be constrained by cosmological arguments) form in the high
energy regime, and the corresponding comoving wave numbers are very large, $k \sim (10^6 - 10^7) k_c$.
We thoroughly studied, by numerical methods, the evolution of scalar perturbation
amplitudes (those needed for calculations of the PBH production) near horizon crossing,
for a wide range of comoving scales. We confirmed the main conclusion of \cite{Cardoso:2007zh}
according to which amplitudes of the curvature perturbation get enhanced after horizon re-entry
(before a beginning of the oscillation phase). We developed an approximate phenomenological approach
for calculations of the perturbation amplitudes for very small scales, where the direct numerical
methods are powerless. We argued, using this approach, that in the asymptotic limit of high energies,
the enhancement factor is constant as a function of the perturbation scale. We presented details 
of the numerical scheme (based on the pseudo-spectral method) which is used for a treating
of scalar cosmological perturbations on the brane and in the bulk.


\end{document}